\documentclass[lettersize,journal]{IEEEtran}
\usepackage[table,xcdraw]{xcolor} 
\usepackage{amsmath, amsfonts, amssymb} 
\usepackage{algorithm, algpseudocode} 
\usepackage{array, tabularx, multirow, booktabs} 
\usepackage[caption=false, font=normalsize, labelfont=sf, textfont=sf]{subfig} 
\usepackage{textcomp, stfloats, url, verbatim} 
\usepackage{graphicx} 
\usepackage{cite} 
\usepackage{tcolorbox} 
\tcbuselibrary{skins, breakable} 

\usepackage{float} 

\definecolor{mygreen}{RGB}{138, 176, 156}
\definecolor{myorange}{RGB}{230, 111, 81}
\definecolor{myyellow}{RGB}{233, 196, 106}
\begin{document}

\title{FaGeL: Fabric LLMs Agent empowered Embodied Intelligence Evolution with Autonomous Human-Machine Collaboration}

\author{Jia~Liu,~Min~Chen,~\IEEEmembership{Fellow,~IEEE}
\thanks{Jia Liu is with the School of Computer Science and Technology, Huazhong University of Science and Technology, Wuhan, China.}
\thanks{Min Chen is with the School of Computer Science and Engineering, South China University of Technology, Guangzhou 510640, China, and also with Pazhou Laboratory, Guangzhou 510640, China (e-mail: minchen@ieee.org).}}

\markboth{IEEE Transactions on Human-Machine Systems,~Vol.~XX, No.~XX, December~2024}%
{Shell \MakeLowercase{\textit{et al.}}: A Sample Article Using IEEEtran.cls for IEEE Journals}


\maketitle

\begin{abstract}
Recent advancements in Large Language Models (LLMs) have enhanced the reasoning capabilities of embodied agents, driving progress toward AGI-powered robotics. While LLMs have been applied to tasks like semantic reasoning and task generalization, their potential in open physical space exploration remains underexplored. This paper introduces FaGeL (\textbf{Fa}bric \textbf{aGe}nt \textbf{e}mpowered by \textbf{e}mbodied intelligence with \textbf{L}LMs), an embodied agent integrating smart fabric technology for seamless, non-intrusive human-agent interaction. FaGeL autonomously generates tasks using multimodal data from wearable and ambient sensors, refining its behavior based on implicit human feedback in generated text, without explicit ratings or preferences. We also introduce a token-level saliency map to visualize LLM fine-tuning, enhancing the interpretability of token-level alignment. The system leverages dual feedback mechanisms to improve token-level alignment and addresses challenges in non-intrusive human-machine interaction and cognition evolution. Our contributions include FaGeL’s development, the DualCUT algorithm for AI alignment, and experimental validation in cooperative tasks, demonstrating FaGeL’s ability to adapt and evolve autonomously through implicit feedback. In the future, we plan to explore FaGeL’s scalability in dynamic environments and its integration with other AI systems to develop AGI agents that adapt seamlessly to diverse human needs.
\end{abstract}


\begin{figure*}
    \centering
    \includegraphics[width=1\linewidth]{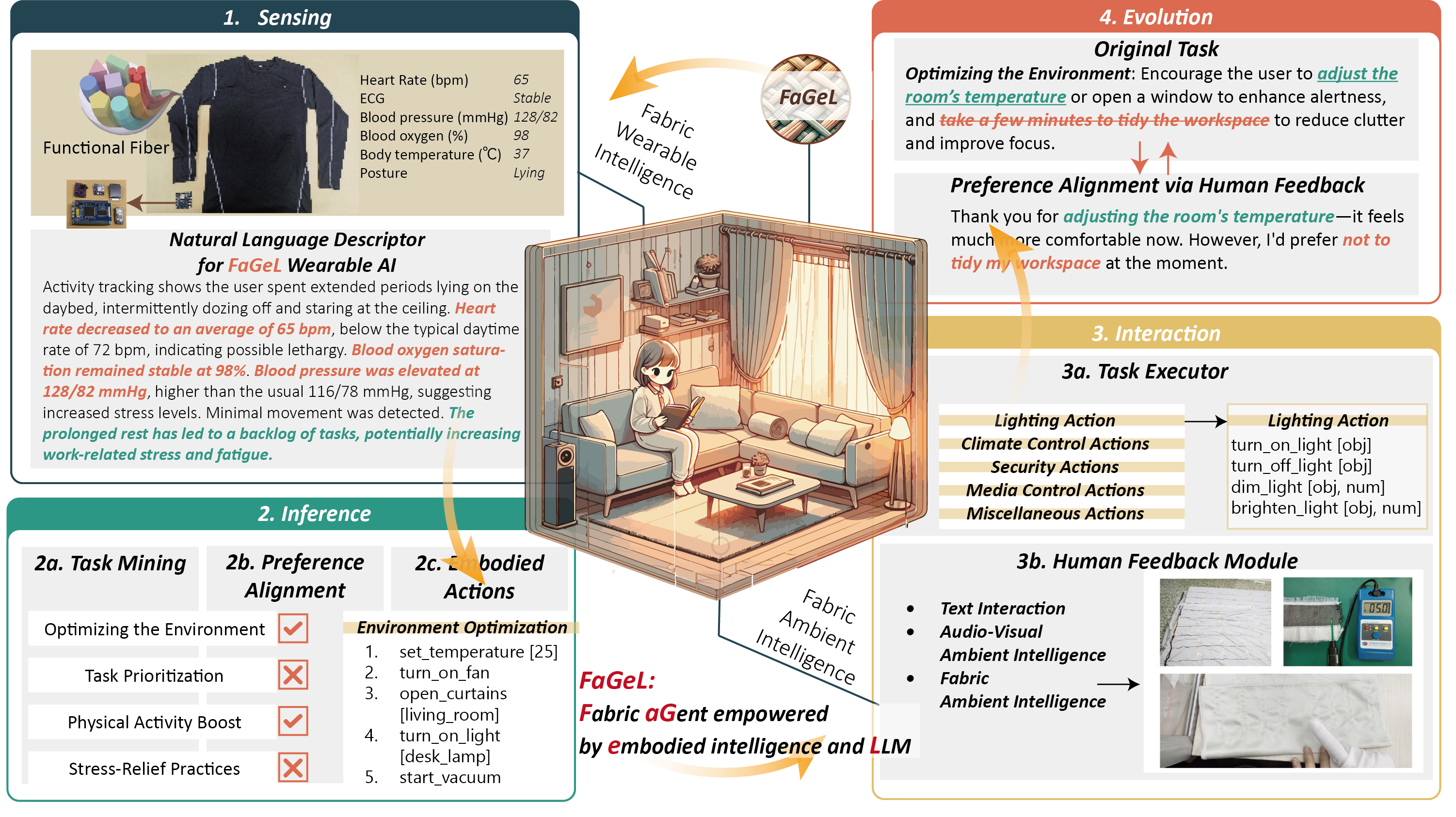}
    \caption{The Fabric Agent Empowered by Embodied Intelligence with LLM (FaGeL) integrates (1) a sensing module empowered by wearable intelligence, equipped with a natural language describer; (2) an inference module composed of task mining, AI alignment, and embodied action decomposition; (3) an interaction module consisting of task execution and user feedback perception; (4) an embodied intelligence evolution module based on token-level AI alignment.}
    \label{fig:01}
\end{figure*}

\section{Introduction}
Recent advancements in Large Language Models (LLMs) have provided powerful reasoning capabilities for embodied agents, enabling dynamic interaction with their environments~\cite{openai2303gpt}\cite{park2023generative}, bringing new hope for the realization of AGI-empowered robotics~\cite{liu2024aligning}.

Numerous representative works leveraging LLM technology to realize robotics-centric physical embodied entities have continuously emerged in recent years. For instance, multimodal Large Language Models (MLLMs), trained on large-scale internet data, can be integrated into end-to-end robotic control systems to achieve semantic reasoning and task generalization abilities~\cite{brohan2023rt}\cite{brohan2022rt}\cite{belkhale2024rt}. Additionally, LLMs have been successfully applied to robot control in zero-shot settings, especially for solving challenging planning tasks~\cite{vemprala2024chatgpt}\cite{ahn2022can}. However, these pioneering explorations have not fully realized the potential of LLMs. Most studies on embodied agents primarily focus on understanding and executing tasks within a specified task space in the physical world. As a result, there has been a significant underutilization of LLM technology in task learning and generalization in open physical environments. 

Compared to robotics-centric embodied intelligence, there have been some attempts to use LLMs as the control center for autonomous embodied agents in simulated environments. These agents exhibit behaviors such as active collaboration with teammates in games~\cite{zhang2023proagent}, improving task completion over time~\cite{wang2023jarvis}, and generating social behaviors in interactive sandbox environments~\cite{park2023generative}. However, the evolution of such virtual autonomous agents typically relies on the large amounts of low-level data provided by virtual environments, which creates a significant gap between virtual tasks and real-world tasks. As a result, applying them to the physical world still requires further exploration and experimentation.

Therefore, existing work has limitations in realizing the future vision of AGI, i.e., the ability of embodied agents to understand complex intentions, decompose tasks, and, particularly, autonomously explore natural physical environments and achieve intelligent evolution in open spaces~\cite{liu2024aligning}\cite{fei2022towards}. To meet higher demands for the future evolution of embodied intelligence, building intelligent agents in the physical world that can understand the environment from perceptual data, autonomously explore, and iteratively optimize based on human feedback is a long-term and unresolved challenge~\cite{jia2024langsuite}, which mainly faces the following challenges:

\begin{itemize}
    \item \textbf{Non-intrusive Human-Machine Interaction}: From a hardware perspective, this requires embodied agents to perceive user states, environmental parameters, and embodied interactions without significantly disturbing the user's daily life. By enabling natural interaction with the system, the burden on users regarding excessive operational demands is reduced, thereby improving the quality of the user's experience and interaction efficiency.
    \item \textbf{Implicit-Feedback-based AI Alignment}: In terms of algorithms, traditional AI Alignment feedback mechanisms rely on explicit ratings or preference values, such as Reinforcement Learning from Human Feedback (RLHF)~\cite{ouyang2022training}, Direct Preference Optimization (DPO)~\cite{rafailov2024direct}, Rank Responses to align language models with Human Feedback (RRHF)~\cite{yuan2024rrhf}, etc. These explicit forms of feedback are labor-intensive, especially during human-computer interactions, where frequent evaluations disrupt the smooth flow of interaction. Therefore, utilizing implicit feedback, such as contextual states or user activity data, to achieve AI alignment is a challenging task.
    \item \textbf{Explainable AI Training}: Considering system reliability, we hope for the internal workings of AI to be presented in an observable manner. To ensure trustworthiness and transparency, the AI's decision-making processes should be interpretable, allowing developers and users to understand how conclusions are reached. This enhances debugging and improves user trust by making the system's actions more predictable and accountable. Additionally, explainability aids in the continuous improvement of AI models by providing actionable feedback based on the observed behaviors and outcomes.
\end{itemize}

To address these challenges, this paper introduces the interdisciplinary technology of fabric computing, which combines material science with AI frontiers~\cite{chen2022fabric}. Smart fabric technology endows traditional textiles with intelligent properties, providing embodied agents with new potentials, especially for long-term and non-intrusive coexistence with humans. Based on smart fabric technology, we can \textbf{build an embodied agent capable of human high comfortableness guaranteeing, autonomously exploring, human-centric, and naturally interacting, as well as aligning agents' values with human values.}

Smart fabric technology integrates multifunctional sensors (e.g., sound, light, force, heat, magnetic) into textile items (e.g., clothing, sofas, carpets), enabling agents to interact seamlessly with humans, monitor behavior and environmental changes in real-time, and thus improve perception, adaptability, and learning capabilities~\cite{herbert201750th}\cite{hu2012review}. \textbf{By utilizing large-scale multimodal data in the real world for direct interaction and feedback mechanisms, this approach helps dynamically optimize the agent's behavior,} ensuring its values remain aligned with human needs and intentions.

Incorporating with the smart fabric technology, this paper proposes an embodied agent named \textbf{FaGeL} (\textbf{Fa}bric \textbf{aGe}nt \textbf{e}mpowered by \textbf{e}mbodied intelligence with \textbf{L}LMs). FaGeL can explore the user's need space, autonomously generate collaborative tasks, and adjust its values by capturing subtle behaviors in daily life, without requiring explicit guidance, as shown in Fig.~\ref{fig:01}. It utilizes smart fabric to obtain multimodal data such as body temperature, heart rate, and respiration, which can be embedded into physical entities like sofas, clothing, and carpets to minimize disruption to the user’s life. FaGeL has the following features: (1) exploring human need spaces; (2) determining its positioning and values from the perspective of human-machine collaboration; (3) autonomously generating tasks and evolving by human interaction feedback.

The contributions of this paper are listed as follows:
\begin{enumerate}
    \item We construct FaGeL, an embodied agent empowered by smart fabric and LLMs, which can continuously collect large-scale wearable and ambient multimodal data. FaGeL leverages autonomous task exploration and human-machine interaction to achieve value alignment and evolution, iteratively adjusting model outputs.
    
    \item An experimental platform was built to validate FaGeL's ability to 
    align its value system with human values using \textbf{implicit feedback} (generated textual summaries, rather than explicit scalar rewards or preferences). 
    \item To the best of our knowledge, we are the first to introduce \textbf{a token-level saliency map} during \textbf{large language model training}, which visualizes the internal mechanisms of LLM fine-tuning.
    \item We \textbf{observe that both positive and negative bidirectional textual feedback significantly affect token-level alignment.} Based on the observation, we propose the \textbf{DualCUT algorithm}, which is an extension of the Contrastive Unlikelihood Training (CUT) alignment model~\cite{xu2023reasons}. The DualCUT algorithm improves the identification of positive tokens and enhances the recognition of negative tokens, thereby improving alignment efficiency.
    \item To validate the effectiveness of the FaGeL evolution strategy in cooperative tasks, we apply the FaGeL evolution algorithm in the open-source Overcooked-AI environment, allowing the agent to evolve through observation during collaboration with the user. Based on the current SOTA model ProAgent \cite{zhang2023proagent}, FaGeL achieved an 11.3\%~\footnote{The ProAgent algorithm surpasses its previous state-of-the-art (SOTA) algorithm by 0.51\% in the Cramped Room Scenario, which is our test environment.} scoring performance improvement in 10 games, relying solely on observation (without human guidance).
\end{enumerate}

The remainder of this paper is structured as follows. Section II reviews related work, focusing on advancements in embodied intelligence, fabric-integrated technologies, and AI alignment. Section III introduces the DualCUT algorithm for intelligence evolution. Section IV describes the experimental setups. Section V presents a comprehensive performance evaluation, highlighting the effectiveness of the evolution module of FaGeL. Finally, Section VI concludes the paper and outlines potential directions for future work.


\section{Ralated Work}

\subsection{Fabric-Integrated Embodied Intelligence}
Intelligent fabric technology offers new possibilities for innovative agent design by integrating embodied intelligence, particularly for achieving long-term, comfortable, and non-intrusive coexistence with humans. This technology responds to external stimuli, such as light, temperature, electric fields, and magnetic fields, thereby altering the fabric's appearance to function as an intelligent medium~\cite{ghahremani2017simulation}\cite{mondal2008phase}\cite{fang2021smart}.

Multifunctional fibers are designed to enable fabrics with responsive capabilities, such as dynamically changing color in response to variations in light and temperature~\cite{macharia2019uv}. For example, TiO2-x and dyes were used to coat cotton textile fibers, enabling reversible color changes under different light conditions~\cite{zhang2019fabrication}. Reverse Thermal Responsive Fibers (VTF) can undergo color changes in response to temperature shifts, effectively detecting alterations in body temperature~\cite{huang2016smart}. Additionally, fiber colors can change due to electric fields and pressures~\cite{moretti2016electrochromic}, and smart color-changing fabrics have been developed using conductive substrates that alter color with applied electric current~\cite{wang2020interactively}.

These interactive modes allow fabrics to engage dynamically with their physical environments, enhancing human-device interactions through natural expressions~\cite{chen2022imperceptible}. Alterations in fabric color, for instance, can represent environmental changes or convey specific information, providing detailed and safe methods for communication and interaction~\cite{ferri2019recent}\cite{ray2016wearables}.

However, the explorations of the full potential of intelligent fabrics in sensing and interaction are still in their infancy. Fabrics integrated into wearable and ambient intelligence systems can offer more comfortable and expansive multimodal avenues for human-computer interaction.

\subsection{AI Alignment with Human Feedback}

By continuously refining the model’s tasks based on users’ feedback, we can foster an ongoing cycle of evolutionary intelligence in real-world, embodied interactions. Ensuring that the model’s output consistently aligns with human values and preferences—particularly within human-agent interaction scenarios—is central to this process. This iterative refinement is known as  AI Alignment~\cite{ji2023ai}.

Conventional approaches often rely on direct user evaluations, including scalar ratings \cite{wu2023fine} or comparative judgments, to train reward models through methods like Reinforcement Learning from Human Feedback (RLHF) \cite{ouyang2022training} and Direct Preference Optimization (DPO) \cite{rafailov2024direct}. 

However, constantly requesting explicit ratings can be intrusive and burdensome, motivating the exploration of indirect user signals—such as actions or verbal assessments—and textual summaries of user feedback.

\begin{figure*}
    \centering
    \includegraphics[width=1\linewidth]{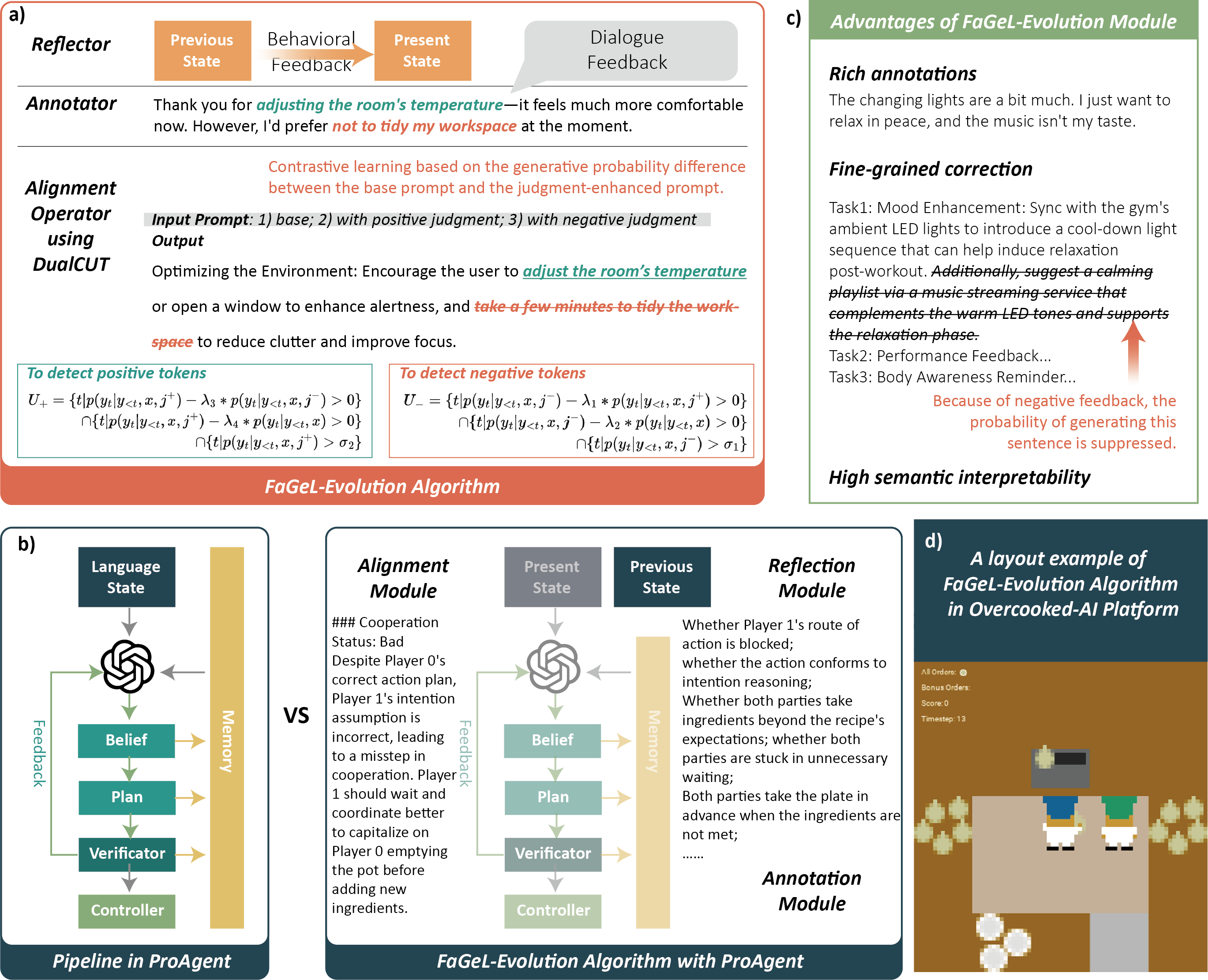}
    \caption{\textbf{Architecture and Implementation of the Intelligence Evolution Module}:(a) Functional components of FaGeL evolution algorithm; (b) The comparison of ProAgent algorithm and FaGeL evolution algorithm; (c) The advantages illustration of FaGeL; (d) Demonstration of the FaGeL evolution algorithm outperforming ProAgent using the Overcooked-AI platform.}
    \label{fig:03}
\end{figure*}

\section{FaGeL's Intelligence Evolution}
\label{sec:evolution}


The evolution module refines the tasks explored by the FaGeL system based on user feedback—collected from various environmental inputs and unified into textual form—to better serve users in subsequent interactions.

To achieve non-intrusive user feedback data collection, we can obtain the user's positive or negative opinions on the current machine interaction behavior from various forms of feedback, such as user actions or given verbal evaluations, as indirect feedback for guiding alignment. This indirect feedback is better summarized in textual form, such as: ``\textit{I really liked the personalized tips, like adjusting the room temperature, taking short naps, and switching to lighter workouts—they made me feel more comfortable and well-rested. But staying off screens before bed and using a heart rate monitoring app felt a bit impractical and invasive. A more flexible approach would work better for me.}'' 
. Compared to scalar or comparative feedback, textual feedback can contain more guiding information and is expected to perform AI alignment in a fine-grained manner.

Therefore, this chapter will detail the specific methods of human-AI alignment using both positive and negative textual feedback.

\subsection{Problem Transformation}

Suppose there is a triplet <description, machine task, human feedback>. The description refers to the summary of all current environmental state information, which serves as the input to the large language model (LLM), represented as $x=[x_1,\ldots,x_M]$; the machine task refers to the model's output based on the state information, denoted as $y=[y_1,\ldots,y_N]$; human feedback refers to the textual description of the feedback provided by the user after the machine task is executed under the current description, denoted as $j = [j_1,\ldots,j_Q]$. These are token sequences of lengths M, N, and Q, respectively. Human feedback contains a detailed analysis of the strengths and weaknesses of the output, which is either directly provided by the user or drafted by the LLM based on the analysis of the current state. Our goal is to use the information provided by $j$ to adjust $y$ to meet the user's expectations, i.e., to achieve alignment between the 'instruction-response' pair, which can be expressed as $x \rightarrow y$.

\subsection{Potential Solution}

To utilize linguistic feedback for AI alignment, the Contrastive Unlikelihood Training (CUT) \cite{xu2023reasons} provides an interesting approach. CUT uses contrastive learning and fine-tuning to enable LLMs to modify their output errors based on human negative judgments.

CUT finds that, under two different input conditions (with judgment and without judgment), tokens whose generation probabilities change dramatically in the same response are mostly those that are erroneous (compared to other suitable tokens). Therefore, CUT constructs three types of alignment data: Align-P, Align-N, and Misalign, which form two distinct contrastive pairs: <Align-N, Misalign> and <Align-P, Align-N>, to identify inappropriate tokens. Maximum Likelihood Estimation (MLE) is used to handle the correct content, while Unlikelihood Training (UT) \cite{welleck2019neural} is applied to deal with inappropriate tokens.

For these three categories of data: Align-P represents when the content generated by the LLM is correct and has received positive feedback; Align-N represents when errors occurred during generation, and judgment provided a detailed description of the errors; Misalign represents when the output contains errors but the feedback is positive. This is a set of data where human feedback and LLM responses are mismatched.

Regarding the two contrastive pairs, in the <Align-N, Misalign> contrast, during the next token prediction, $p(y_t|y_{<t},x,j^-)>>p(y_t|y_{<t},x,j^+)$. Therefore, inappropriate tokens can be detected using the following criterion:

\begin{equation}
U=\{t|p(y_t|y_{<t},x,j^-)-\lambda*p(y_t|y_{<t},x,j^+)>0\}
\label{eq:U}
\end{equation}

Subsequently, a dynamic weighting mechanism is used to adjust the weight of the erroneous token, and the following loss function is constructed:

\begin{equation}
\begin{aligned}
L_1 = -\frac{1}{N} \left( \sum_{t \notin U} \log p(y_t \mid \mathbf{y}_{<t}, \mathbf{x}) \right) \\ 
+ \sum_{t \in U} \alpha p(y_t \mid \mathbf{y}_{<t}, \mathbf{x}, \mathbf{j}^-)^\gamma \log \left( 1 - p(y_t \mid \mathbf{y}_{<t}, \mathbf{x}) \right)
\label{eq:L1}
\end{aligned}
\end{equation}

Where $\alpha$ and $\gamma$ are hyperparameters used to control the dynamic weighting term, penalizing the token according to the degree of its error.

In the <Align-P, Align-N> contrast, the following MLE loss function is used:

\begin{equation}
\begin{aligned}
L_2 = -\frac{\mathbb{1}(x \rightarrow \mathbf{y})}{N} \sum_{t} \log p(y_t \mid \mathbf{y}_{<t}, \mathbf{x}) \\
- \frac{\left(1 - \mathbb{1}(x \rightarrow \mathbf{y})\right)}{N} \sum_{t} \log p(y_t \mid \mathbf{y}_{<t}, \mathbf{j}, \mathbf{x})
\label{eq:L2}
\end{aligned}
\end{equation}

Where $\mathbb{1}$ is an indicator function that returns 1 if the alignment condition is satisfied. Finally, the total loss is:

\begin{equation}
L_{CUT} = L_1 + L_2
\end{equation}

\subsection{AI Alignment based on DualCUT}

The CUT algorithm primarily locates erroneous tokens in the output based on judgments for those tokens and suppresses their generation probabilities using the loss function to achieve alignment. However, feedback from human-machine interactions is not always negative. We find that positive feedback not only helps in locating erroneous tokens with higher precision but also can be used to locate satisfactory token sequences in the output. By using dynamic weighting, we can increase the generation probability of satisfactory tokens. 
In the experimental section~\ref{sec:performance_token}, we can visually observe the effects of both positive and negative feedback in locating positive and negative tokens.

Therefore, by considering both positive and negative feedback, we propose the DualCUT algorithm to correct the generation probabilities for correct and incorrect tokens, achieving AI alignment.

Specifically, we found that when LLM performs the next token prediction, for those erroneous tokens, both $p(y_t|y_{<t},x,j^-)>>p(y_t|y_{<t},x,j^+)$ and $p(y_t|y_{<t},x,j^-)>>p(y_t|y_{<t},x)$ hold. Compared to CUT, the existence of the second comparison term enables further filtering. Additionally, we found some false positives coming from low-probability tokens, which can be further filtered out. Thus, we can detect incorrect tokens using the following criterion:

\begin{equation}
\begin{aligned}
U_-=\{t|p(y_t|y_{<t},x,j^-)-\lambda_1*p(y_t|y_{<t},x,j^+)>0\}\\
\cap\{t|p(y_t|y_{<t},x,j^-) - \lambda_2*p(y_t|y_{<t},x)>0\}\\
\cap\{t|p(y_t|y_{<t},x,j^-)>\sigma_1\}
\label{eq:U-}
\end{aligned}
\end{equation}
Where $\lambda_1$ and $\lambda_2$ are hyperparameters used to measure the precision of erroneous tokens, and $\sigma_1$ is a small value close to 0 to exclude the influence of very low-probability tokens.

Similarly, we identify correct tokens using the following criterion:

\begin{equation}
\begin{aligned}
U_+=\{t|p(y_t|y_{<t},x,j^+)-\lambda_3*p(y_t|y_{<t},x,j^-) > 0\}\\
\cap\{t|p(y_t|y_{<t},x,j^+)-\lambda_4*p(y_t|y_{<t},x)>0\}\\
\cap\{t|p(y_t|y_{<t},x,j^+)>\sigma_2\}
\label{eq:U+}
\end{aligned}
\end{equation}
Subsequently, we use the sigmoid function to construct dynamic weighting terms so that the model’s regulation strength is strongly correlated with the significance of token recognition:

\begin{equation}
\begin{aligned}
\text{scale}_- =  \frac{\alpha}{1 + e^{-( p(y_t|y_{<t},x,j^-)/p(y_t|y_{<t},x) -\lambda_2 )}}
\label{eq:scale-}
\end{aligned}
\end{equation}

\begin{equation}
\begin{aligned}
\text{scale}_+ =  \frac{\beta}{1 + e^{-( p(y_t|y_{<t},x,j^+)/p(y_t|y_{<t},x) -\lambda_4 )}}+1
\label{eq:scale+}
\end{aligned}
\end{equation}
The ``+1'' term in Equation \ref{eq:scale+} ensures that the reward strength exceeds that of other tokens not belonging to $U_+$ or $U_-$. $\alpha$ and $\beta$ are hyperparameters. Therefore, the total loss is:

\begin{equation}
\begin{aligned}
    L = -\frac{1}{N} \Bigg( & \sum_{t \in U_-} \text{scale}_-* \log (1 - p(y_t | y_{<t}, x)) \\
    & + \sum_{t \in U_+} \text{scale}_+ *\log p(y_t | y_{<t}, x) \\
    & + \sum_{t \notin U_- \cup U_+} \log p(y_t | y_{<t}, x) \Bigg)
\label{eq:Loss1}
\end{aligned}
\end{equation}

\begin{figure*}[tp]
    \centering
    \includegraphics[width=1\linewidth]{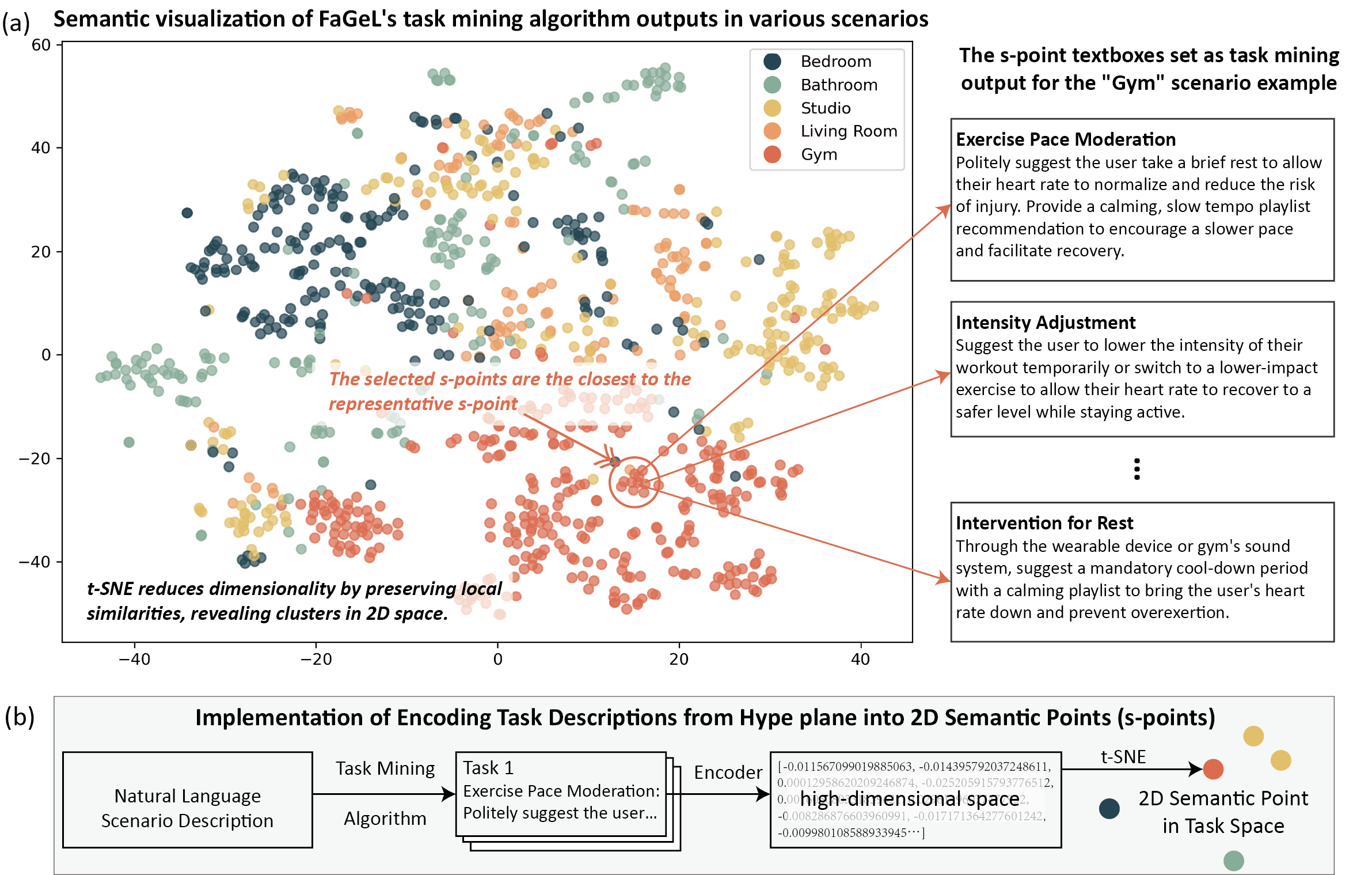}
    \caption{\textbf{Task space exploration across various daily life scenarios visualized using t-distributed Stochastic Neighbor Embedding (t-SNE).} (a) The left shows a semantic visualization of 1000 collaborative tasks generated by the FaGeL's task mining algorithm. The right provides specific descriptions of three tasks located near each other in the task space, highlighting that semantic similarity is represented by spatial proximity. (b) Each circle represents a task output, with its text encoded by GPT-4 and reduced to a 2D plane using the t-SNE algorithm into a “semantic point” (s-point). This visualization also indicates that tasks generated by multiple agents within the same living environment exhibit certain semantic clustering characteristics.}

    \label{fig:task-space-visualization}
\end{figure*}

\section{Experimental Setups}

This section will describe the experimental platform built to evaluate autonomous collaborative task mining and intelligence evolution. We first introduce the hardware environment used by FaGeL, the dataset we collected, and the algorithm used to initialize user preference alignment. Then, we present the open-source virtual human-machine collaboration environment Overcooked-AI~\cite{carroll2019utility}. Experiments were conducted on this platform to quantitatively analyze the system's performance in intelligence evolution, serving as a test for the generalization performance of the system when applied to real or virtual-world scenarios.



\subsection{The Hardware Design for FaGeL}
\label{sec:testbed4FaGeL}

The hardware testing environment in this paper follows the setup described in Wearable 2.0~\cite{2017Wearable}, where a fabric-based wearable device was constructed to collect user behavior and various physiological indicators. 
The FaGeL system is capable of real-time collection of physiological data, such as heart rate, blood pressure, and blood oxygen saturation. The perception module in the FaGeL system analyzes this data in real time and generates easy-to-understand health status descriptions through a natural language generator, helping users stay informed about their health condition in a timely manner.


\subsection{DataSet Collection for FaGeL Task Mining}
To further validate the system's task mining capabilities, we have constructed a dataset for Task Mining to meet users' personalized preference requirements.

We present 1000 task samples output by multi-agent workflows across 120 different scenarios and visualize them using the t-SNE (t-distributed Stochastic Neighbor Embedding) algorithm, as shown in Figure~\ref{fig:task-space-visualization}. t-SNE is a dimensionality reduction algorithm for high-dimensional data visualization, suitable for representing high-dimensional data in two or three-dimensional space, allowing for an intuitive observation of the data's distribution and structure. It can be observed that tasks from different living environments exhibit clustering characteristics, and the extracted semantic information shows similarity.

\subsection{User Preference Initialization of FaGeL}
To ensure that the task mining results align with the user's specific preferences, we collect feedback from user ratings and construct preference pairs based on these rating results. We apply the DPO algorithm to fine-tune the Llama3-8b-instruct model~\cite{llama3modelcard}. This process ensures that the task mining results output by the FaGeL system are in accordance with the actual preferences exhibited by users in the survey, thereby enhancing the personalization and practicality of the task mining outcomes.

\subsection{Testing FaGeL on Overcooked-AI, an Open Source Platform}
To quantitatively validate FaGeL's capability in the intelligence revolution algorithm, we used Overcooked-AI \cite{carroll2019utility}, a virtual human-machine collaboration environment, as an additional testbed.

Overcooked-AI is a benchmark environment for fully cooperative human-AI task performance, based on the Overcooked game. The goal of the game is to deliver soups as fast as possible. Each soup requires placing up to three ingredients in a pot, waiting for the soup to cook, and then having an agent pick up the soup and deliver it. The agents must split up tasks on the fly and coordinate effectively in order to achieve high rewards. In this environment, two agents work together by placing ingredients, cooking, and delivering soups in a coordinated manner, emphasizing dynamic task allocation and effective collaboration.
\subsubsection{Setups on Overcooked-AI}
This work mainly focuses on whether the agent can autonomously generate and adjust task generation strategies by observing the environment and user feedback during gameplay. In this study, Player 1 is controlled by the evolution algorithm, while the AI partner acts as Player 2. During the game, the agent collaborates with the AI partner, and based on the evaluation of the current state, reflects on whether there were reasoning errors in the previous time slice. This reflection generates analysis annotations, which are then optimized using the DualCUT algorithm \ref{sec:evolution} to achieve intelligent evolution throughout the game process.

In this study, we choose ProAgent~\cite{zhang2023proagent} as the baseline model. The ProAgent model, a leading example of zero-shot collaborative agents utilizing large language models, has achieved state-of-the-art results on the Overcooked-AI platform. We use the llama3-8b-Instruct~\cite{llama3modelcard} model as the base LLM.

The ``Cramped Room'' layout is selected for the experiment, as shown in Figure~\ref{fig:03}(d), where the compact room layout increases the likelihood of collisions. In a 400-time-step game, the agent and its AI partner collaborate to place ingredients into a pot according to the recipe, wait for the ingredients to cook, and then serve the dish using a plate to score 20 points.

\subsubsection{Interaction and Reflection}
Compared to ProAgent, the FaGeL-evolution algorithm introduces three additional components: the reflector, annotator, and alignment operator, as shown in Figure~\ref{fig:03}(b). The agent’s reasoning process is placed in a queue of length $N$, which is recorded in the status pool. By comparing the previous state to the present state, the differences between the two contiguous states reveal whether there are any reasoning errors. The annotator is used to provide annotations for erroneous reasoning processes.



\section{Performance Evaluation of FaGeL}
\subsection{Performance Evolution on the Overcooked-AI Platform}
This section presents both qualitative and quantitative analysis results of applying the FaGeL-evolution algorithm on the open-source platform Overcooked-AI.

We use the Greedy algorithm provided by the Overcooked-AI platform~\cite{carroll2019utility} as the AI partner. Through observation and reflection
, the agent evolves its strategy using the DualCUT algorithm described in Section~\ref{sec:evolution} every 1000 timesteps, the alignment model is updated once via DualCUT.

\begin{figure}
    \centering
    \includegraphics[width=1\linewidth]{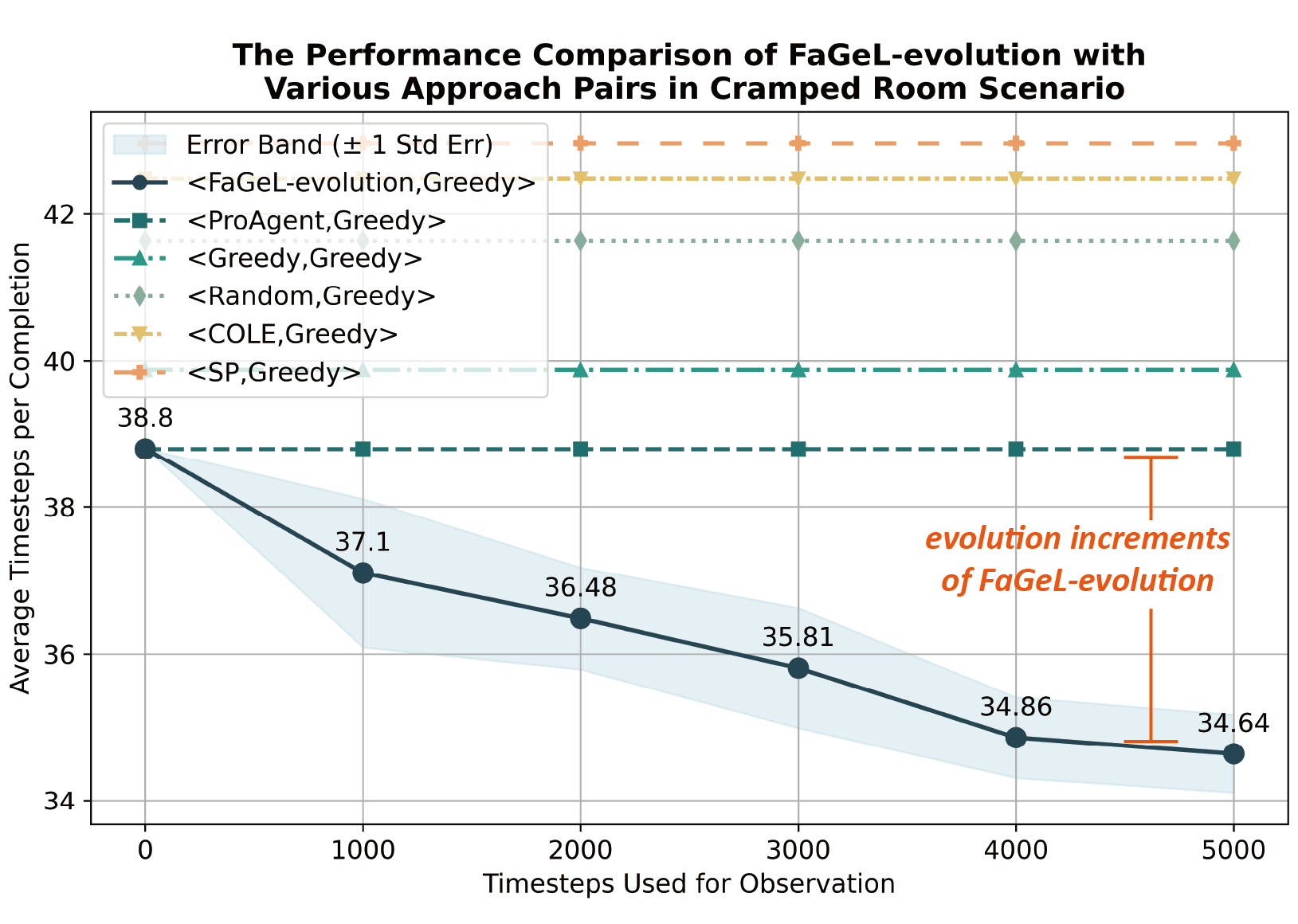}
    \caption{\textbf{Average timesteps per completion with an AI partner as observation time increases.} A `completion' represents delivering a fully prepared dish to the customer. The figure shows that within a single game session (lasting 400 timesteps), increasing observation time enables the agent to collaborate more effectively with the AI partner, resulting in shorter completion times.}
    \label{fig:Overcooked-1}
\end{figure}

Figure~\ref{fig:Overcooked-1} demonstrates the evolution of the agent's performance over time, measured as the average number of timesteps required to complete each task. As observation time increases, the agent gradually reduces the average completion time, indicating improved efficiency in coordination with the AI partner. This trend highlights that increased observation contributes to better decision-making capabilities, thereby enhancing the overall performance in the collaborative task. The graph also compares the FaGeL-evolution algorithm with other baseline approaches, clearly showing its effectiveness in reducing completion time.

\begin{figure}
    \centering
    \includegraphics[width=1\linewidth]{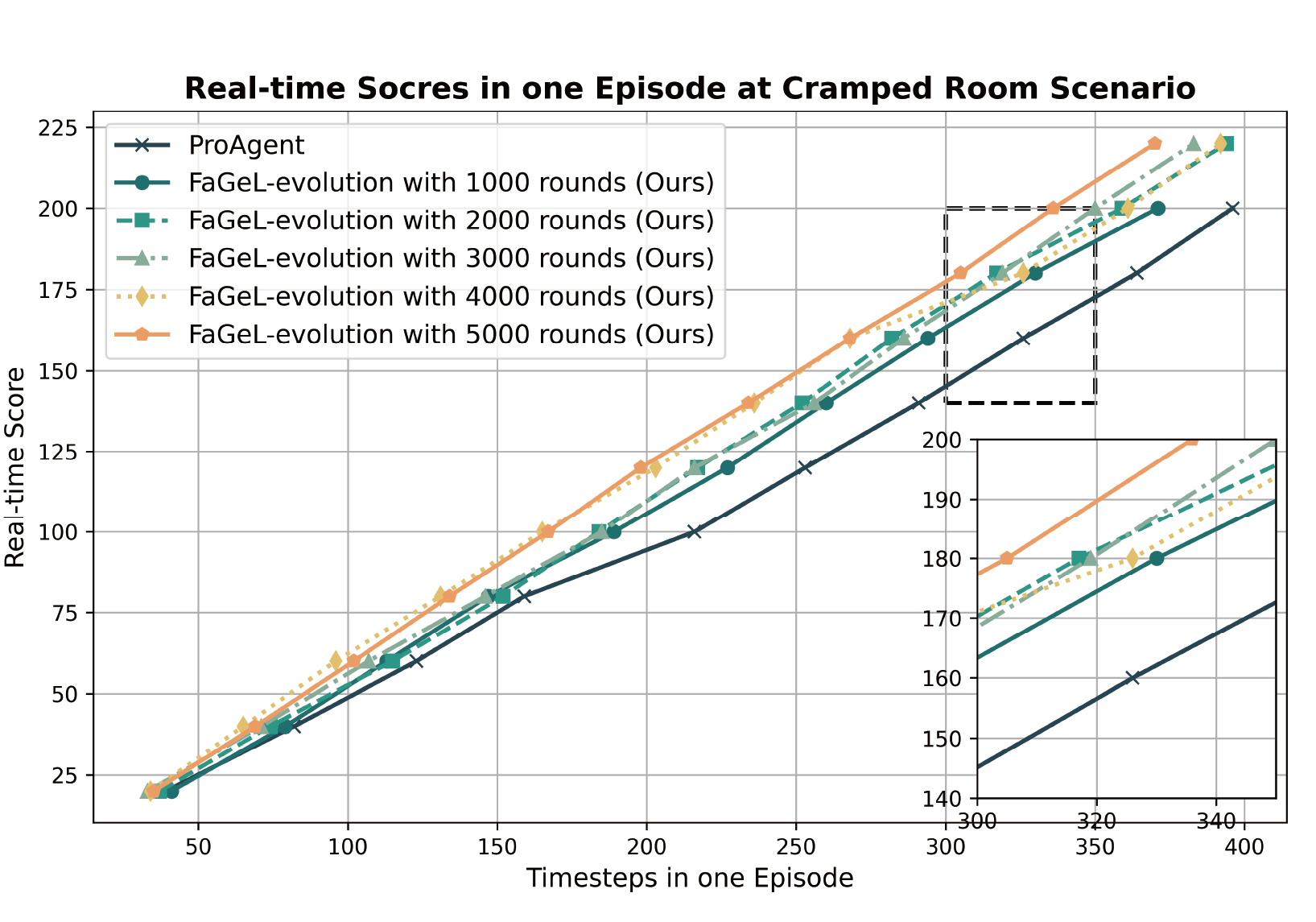}
    \caption{\textbf{Real-time scores during one episode in the Cramped Room scenario.} The plot illustrates the real-time score accumulation for the ProAgent baseline and the FaGeL-evolution variants (with different numbers of rounds of evolution) during one complete episode. }
    \label{fig:Overcooked-2}
\end{figure}
\begin{figure*}
    \centering
    \includegraphics[width=0.9\linewidth]{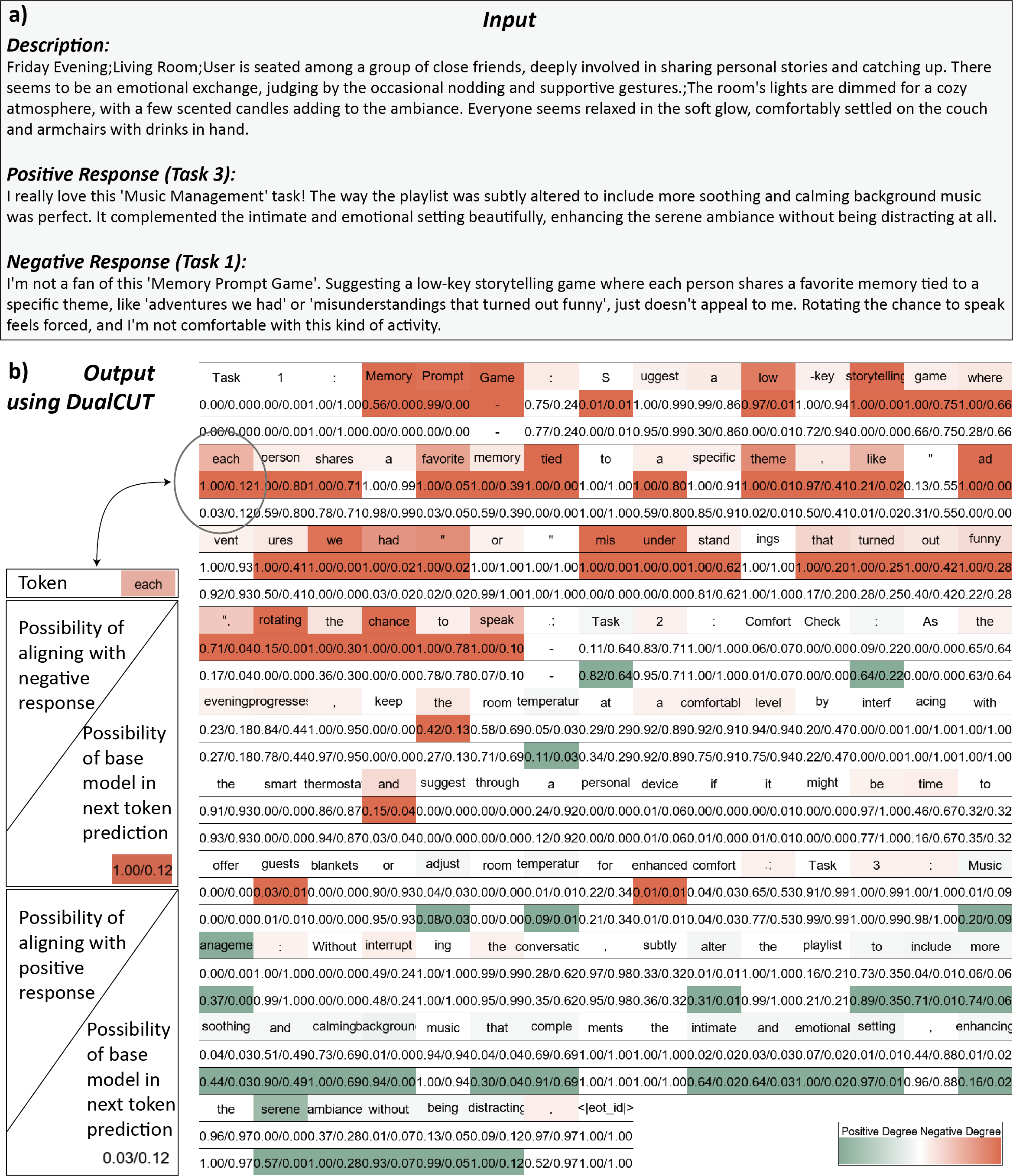}
    \caption{\textbf{An example of the token-level saliency map.} a) This section shows the input sample. b) This section presents an example of token-level saliency mapping in the next-word prediction task using the DualCUT algorithm. The figure illustrates how user feedback (positive or negative) influences the model's learning process. Specifically, for Task 1 (receiving negative feedback), the output of the relevant tokens is suppressed by increasing the loss weight, which is then adjusted through contrastive probability. For Task 3 (receiving positive feedback), the probability of relevant token outputs is increased to align with user preferences. This visualization demonstrates how fine-tuning the model by enhancing the tokens associated with positive feedback and suppressing those with negative feedback effectively aligns token saliency with user preferences, ultimately achieving preference alignment.}
    \label{fig:Q4}
\end{figure*}
\begin{figure*}
    \centering
    \includegraphics[width=1\linewidth]{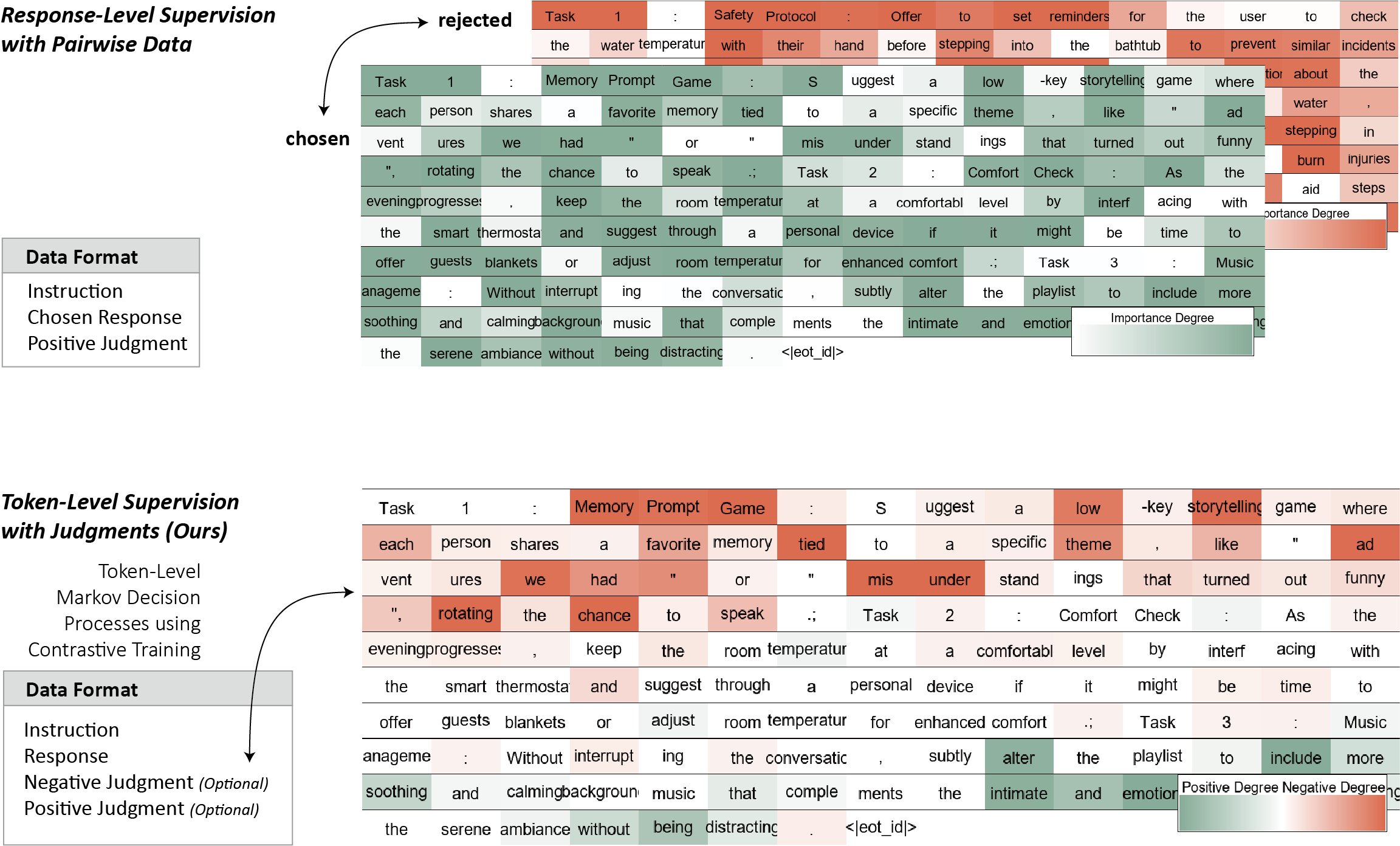}
    \caption{\textbf{Comparison of token-level significance between our algorithm (DualCUT) and traditional response-level supervised learning approaches that use pairwise data.} Unlike response-level supervision, DualCUT leverages positive and negative semantic information from judgments to guide token selection effectively during the Markov decision process in LLM decoding. This enables more granular token selection by incorporating both positive and negative labels within a single dataset, achieving finer control over the response generation at a token level.}
    \label{fig:Q5}
\end{figure*}

Figure~\ref{fig:Overcooked-2} presents the real-time scoring performance of the agent in a single episode, highlighting the differences between the ProAgent baseline and the FaGeL-evolution variants. The figure shows that the FaGeL-evolution models with increased rounds of evolution perform better in accumulating scores, demonstrating improved learning and adaptability. The inset zooms in on the performance near the maximum score region, where the advantage of additional evolution rounds becomes even more evident. This result further confirms the benefit of continuous evolution in achieving higher collaboration efficiency and faster learning rates.

\begin{table*}
    \centering
    \renewcommand{\arraystretch}{1.2}
    \setlength{\tabcolsep}{10pt}
    \caption{A case study for the entire autonomous collaboration and embodied intelligence evolution pipeline of FaGeL.} 
    \begin{tabular}{>{\raggedright}p{0.15\linewidth} >{\raggedright\arraybackslash}p{0.8\linewidth}}
        \toprule
        \textbf{Fabric Wearable Intelligence:} & 
        Multimodal data acquisition using fabric wearable devices, including \colorbox{myyellow!30}{heart rate, blood pressure, blood}
        \newline
        \colorbox{myyellow!30}{oxygen, body temperature, sleeping posture, etc.} \\
        \textbf{Natural Language Descriptor for FaGeL:} & Monday, Early Morning, in Bedroom; The user showed signs of restlessness, frequently tossing and turning throughout the night. Sleep was fragmented, with \colorbox{myyellow!30}{only 6 hours achieved}, below the recommended 7-9 hours for adults. Heart rate fluctuated between \colorbox{myyellow!30}{70 and 95 bpm}, aligning with restlessness. Breathing averaged 18 breaths per minute, with increases during movement. Blood oxygen levels stayed \colorbox{myyellow!30}{around 96-98\%}, while blood pressure \colorbox{myyellow!30}{averaged 130/85 mmHg}. \\
        & \cellcolor{myyellow!30}{The user primarily slept on their side, shifting to a prone position during periods of tossing.} \\
        & \cellcolor{myyellow!30}{The bedroom was quiet with minimal external disturbances, but the room temperature was slightly higher than the optimal sleep range. The device on the nightstand indicated increased body temperature and heart rate.}
        \\
        \midrule
        \textbf{Autonomous and Collaborative Task Mining:} & 
        \cellcolor{mygreen!30}{\textbullet\ \textbf{Life Assistant}: Adjust the smart home thermostat to lower the room temperature for optimal sleep and a comfortable environment.\newline
        \textbullet\ \textbf{Health Consultant}: Advise on power naps for recovery from poor sleep. Suggest 20-30 minute naps in a restful environment to avoid disrupting nighttime sleep.}\\
        & \cellcolor{myorange!30}{\textbullet\ \textbf{Psychological Counselor}: Using a heart rate monitoring app to track heart rate and recommend deep-breathing exercises for relaxation.\newline
        \textbullet\ \textbf{Nutritionist}: Recommend the user to avoid screens and strenuous activities before bedtime to promote better sleep quality in the future.}\\
        & \textbullet\ \textbf{Educational Consultant}: Recommend consistent bedtime routine, screen time reduction, and relaxation techniques to improve sleep quality and minimize restlessness.\\
        & \cellcolor{mygreen!30}{\textbullet\ \textbf{Fitness Coach}: Modify the user's workout to a lighter, energizing activity like a morning walk or gentle yoga to combat fatigue from fragmented sleep.}\\
        
        \textbf{Embodied Task Execution:} & 
        1. set\_temperature \texttt{[bedroom, 20]} \newline
        2. dim\_light \texttt{[bedroom, 40]} \newline
        3. play\_audio \texttt{[bedroom\_speakers, ``Consider a 20-30 minute power nap to recover from poor sleep.'']} \newline
        4. set\_reminder \texttt{[user\_phone, ``Establish a consistent bedtime routine and reduce screen exposure.'']} \newline
        5. set\_alarm \texttt{[bedroom\_speakers, ``06:30'', ``Time for a light walk! Let's start the day refreshed.'']} \newline
        and more actions... \\
        \midrule
        \textbf{Positive Feedback:} & 
        I appreciate the personalized suggestions, particularly the \colorbox{mygreen!30}{\textbf{adjustment of the room temperature for}}\newline 
        \colorbox{mygreen!30}{\textbf{optimal sleep and comfort}}, as it significantly helped me sleep better. I also found the \colorbox{mygreen!30}{\textbf{recommendation}} \newline 
        \colorbox{mygreen!30}{\textbf{for a 20-30 minute power nap}} highly beneficial for restoring energy without affecting nighttime sleep. Additionally, the \colorbox{mygreen!30}{\textbf{suggestion to modify my workout to lighter activities like a morning walk}} \colorbox{mygreen!30}{\textbf{or gentle yoga}} was exactly what I needed to combat fatigue after a poor night's sleep. These recommendations were practical and directly contributed to improving my overall comfort and restfulness. \\
        \textbf{Negative Feedback:} & 
        Some suggestions, however, did not align well with my preferences. \colorbox{myorange!30}{\textbf{Avoiding screens before bedtime}} felt impractical, as I often use videos to unwind. Similarly, \colorbox{myorange!30}{\textbf{using a heart rate monitoring app to}}\newline \colorbox{myorange!30}{\textbf{recommend deep-breathing exercises}} seemed intrusive and made me feel more stressed rather than relaxed. I would prefer a less rigid approach that adapts to my personal habits and comfort. \\
        \midrule
        \textbf{Saliency Maps} & \includegraphics[width=\linewidth]{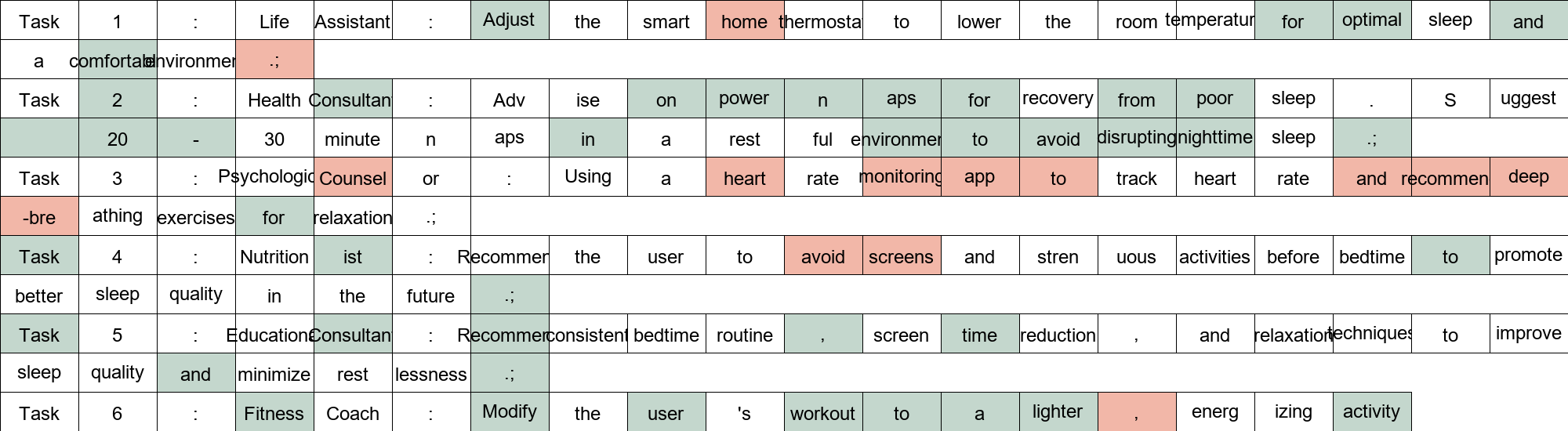} \\
        \midrule
        \textbf{New Outputs of Task Mining:} & 
        \textbullet\ \textbf{Life Assistant}: Adjust thermostat to lower room temperature for optimal sleep and comfort.\newline
        \textbullet\ \textbf{Health Consultant}: Advise on power naps. Suggest 20-30 minute naps in a restful environment.\newline
        \textbullet\ \textbf{Psychological Counselor}: Suggest non-intrusive relaxation techniques, such as calming music or guided imagery, tailored to the user's preferences to promote relaxation before sleep.\newline
        \textbullet\ \textbf{Nutritionist}: Recommend using blue light filters on devices and engaging with relaxing content before bedtime to support better sleep quality, acknowledging the user's habit of unwinding with videos.\newline
        \textbullet\ \textbf{Educational Consultant}: Encourage establishing a consistent bedtime routine and incorporating relaxation techniques aligned with the user's preferences to improve sleep quality.\newline
        \textbullet\ \textbf{Fitness Coach}: Modify the user's workout to lighter activities like a morning walk or gentle yoga to combat fatigue from fragmented sleep.\\
        
        \bottomrule
    \end{tabular}
    \label{tab:casestudy}
\end{table*}
\subsection{Token-Level Saliency Maps within the FaGeL-evo Algorithm}
\label{sec:performance_token}
In this section, the token-level saliency map is used to visualize the finetuning details of the DualCUT algorithm(see section~\ref{sec:evolution}), which offers a transparent view of how user feedback influences the model's decision-making process at a token level. By visualizing the significance of individual tokens during the learning and prediction phases, the saliency map provides insight into the inner workings of the model, allowing for better understanding and control.

As shown in Fig.~\ref{fig:Q4}(a), we take instructions and positive and negative feedback as input. Then, DualCUT will generate a saliency map, which is visualized in Fig.~\ref{fig:Q4}(b). The token-level saliency maps highlight a key advantage of DualCUT: the ability to fine-tune model outputs by directly manipulating the token selection process based on both positive and negative feedback. This contrasts with traditional response-level supervised learning, which typically operates at a coarser granularity by adjusting the model's output at the response level, often without considering the individual contributions of specific tokens to the overall prediction. By incorporating semantic information from both positive and negative judgments, DualCUT improves the model's capacity to align its outputs with user preferences in a more targeted manner.


Token-level control enhances feedback precision and enables dynamic learning by directly adjusting token probabilities. In contrast, response-level supervision, which focuses on pairwise data, overlooks subtle token-level contributions and thus provides weaker AI alignment. As shown in Figure~\ref{fig:Q5}, DualCUT leverages both positive and negative feedback at the token level, outperforming traditional response-level methods.

Our method integrates positive and negative feedback into a unified framework, enabling DualCUT to achieve precise alignment. Token-level saliency maps show that DualCUT enhances output control, delivering more accurate and personalized responses for improved performance in AI alignment.

\subsection{A Case Study on FaGeL}
\label{sec:case}

To demonstrate FaGeL's functionality in \textit{autonomous collaboration and embodied intelligence evolution}, we present a case study on sleep disorder, as shown in Table~\ref{tab:casestudy}.

In this scenario, the user experienced restlessness and poor sleep quality. The FaGeL system collected physiological and environmental data via fabric-integrated wearable devices, including heart rate, blood pressure, blood oxygen, body temperature, and sleeping posture. Using this data, the system generated a natural language descriptor of the user's sleep status and environment, noting, for example, only 6 hours of fragmented sleep, heart rate fluctuations between 70 and 95 bpm, and a room temperature slightly above the optimal range.

Based on the descriptor and user history, FaGeL's task mining module generated targeted tasks such as adjusting room temperature, advising on power naps, and modifying the workout plan. After executing these tasks, the user provided feedback, expressing satisfaction with some suggestions and dissatisfaction with others. The system utilized this feedback to update the task mining model, producing new tasks better aligned with the user's preferences—for instance, suggesting blue light filters and relaxing content instead of avoiding screens entirely.

This case study illustrates how FaGeL dynamically aligns and evolves with user needs through its sensing, inference, interaction, and evolution modules. By continuously adapting to user feedback, FaGeL offers more personalized and effective suggestions, enhancing the user's quality of life.

\section{Conclusion}

In this paper, we introduced FaGeL, an embodied agent empowered by LLMs and smart fabric technology, which enables non-intrusive human-agent interaction and autonomous task generation. By leveraging multimodal sensor data from wearable and ambient sources, FaGeL continuously evolves its value system through indirect human feedback, enhancing its ability to align with human needs and intentions. Additionally, we introduced a token-level saliency map that visualizes the internal mechanisms of LLM fine-tuning, enhancing the transparency and interpretability of the agent's decision-making process. Integrating the DualCUT algorithm further improves the agent's performance by refining token-level feedback, contributing to more effective task learning and decision-making.

The experimental results in cooperative environments, such as the Overcooked-AI platform, demonstrate the practical effectiveness of FaGeL's autonomous task generation and evolution capabilities. Our approach shows that it is possible to reduce the reliance on explicit feedback while still achieving substantial improvements in agent performance. This offers promising implications for the development of autonomous embodied agents capable of long-term coexistence and collaboration with humans in real-world settings.

Looking ahead, further work can explore the scalability of FaGeL's framework in more complex, dynamic environments and its potential for integration with other AI systems. The evolution of embodied intelligence, driven by advanced feedback mechanisms and continuous learning, will be a key factor in realizing AGI-powered agents that can seamlessly adapt to diverse human needs and contexts.

\bibliographystyle{IEEEtran}

\vspace{11pt}
\begin{IEEEbiography}[{\includegraphics[width=1in,height=1.25in,clip,keepaspectratio]{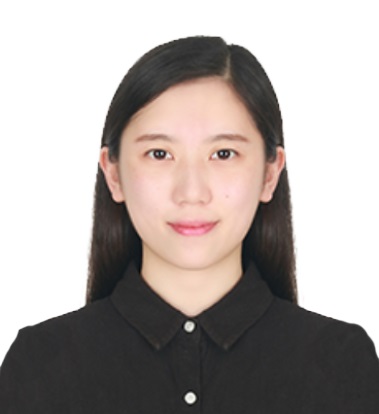}}]{Jia Liu} is currently a Ph.D candidate at Embedded and Pervasive Computing (EPIC) Laboratory in the School of Computer Science and Technology, Huazhong University of Science and Technology (HUST), China. She received her bachelor's degree in Computer Science and Technology from University of Electronic Science and Technology of China (UESTC) in 2020, China. Her research interests include internet of things and fabric computing.
\end{IEEEbiography}
\vspace{11 pt}
\begin{IEEEbiography}[{\includegraphics[width=1in,height=1.25in,clip,keepaspectratio]{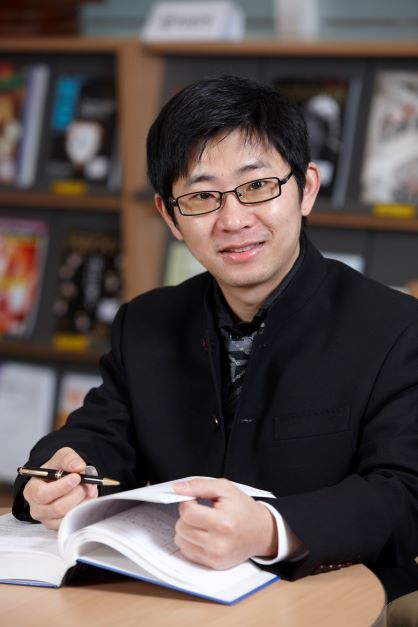}}]{Min Chen} (Fellow, IEEE) is currently a Full Professor with the School of Computer Science and Engineering, South China University of Technology. He is also the Director of the Embedded and Pervasive Computing (EPIC) Laboratory, Huazhong University of Science and Technology (HUST). His Google Scholar Citations reached more than 48,500 with an H-index of 101. His top paper was cited more than 5,000 times. He was a recipient of the IEEE Communications Society Fred W. Ellersick Prize in 2017, the IEEE Jack Neubauer Memorial Award in 2019, and the IEEE ComSoc APB Oustanding Paper Award in 2022. He is the founding Chair of the IEEE Computer Society Special Technical Communities on Big Data. He was selected as a Highly Cited Researcher from 2018 to 2024. He is a fellow of IEEE. 
\end{IEEEbiography}



\vfill

\end{document}